\documentclass{aa}
\usepackage{graphicx}
\usepackage{multirow}
\usepackage{txfonts}
\usepackage{floatflt}
\usepackage[authoryear]{natbib}
\bibpunct{(}{)}{;}{a}{}{,}

\begin{document}

   \title{Evolutionary history of four binary blue stragglers from the
    globular clusters $\omega$ Cen, M55, 47 Tuc and NGC 6752}

   \author{K. St\c{e}pie\'n, \inst{1}
           A. A. Pamyatnykh, \inst{2}
           \and
           M. Rozyczka \inst{2}
           }

   \institute{Warsaw University Observatory,
              Al. Ujazdowskie 4, PL-00-478 Warsaw\\
              \email{kst@astrouw.edu.pl}
         \and
              Nicolaus Copernicus Astronomical Center,
              Bartycka 18, PL-00-716 Warsaw\\
              \email{[alosza, mnr]@camk.edu.pl}
             }

   \date{Received ; accepted }

 
  \abstract
{Origin and evolution of blue stragglers in globular clusters is still
a matter of debate.}
{The aim of the present investigation is to reproduce the evolutionary
  history of four binary blue stragglers in four different clusters, 
  for which precise values of global parameters are known.}
{Using the model for cool close binary evolution, developed by one of
  us (KS), progenitors of all investigated binaries were found and 
  their parameters evolved into the presently observed values.}
 {The results show that the progenitors of the binary blue stragglers
   are cool close binaries with period of a few days, which transform 
   into stragglers by
   rejuvenation of the initially less massive component by mass
   transfer from its more massive companion overflowing the inner
   critical Roche surface.

   The parameters of V209 from ${\omega}$ Cen indicate that the binary
   is substantially enriched in helium. This is an independent and strong
   evidence for the existence of the helium rich subpopulation in this
   cluster.}
{}


   \titlerunning{Evolutionary history of four binary blue stragglers}
   \authorrunning{K. St\c{e}pie\'n et al.}
   \maketitle
%

\section{Introduction}
\label{sect:intro}

Blue stragglers (BSs) are usually defined as objects brighter and bluer than the main-sequence 
turn-off point of their host cluster \citep{sand53}. In the color-magnitude diagram (CMD) they 
occupy an extension of the main sequence (MS), from which all normal stars at the age of the cluster 
would have evolved away billions of years ago. In other words, they indeed seem to lag in time, or 
straggle, behind the bulk of the cluster \citep{sha97}. Their location in the CMD suggests typical 
masses of 1.0-1.5 $M_\odot$ - significantly larger than those of normal stars in globular 
clusters (GCs). Because of that, BSs are thought to have increased their mass during their 
evolution. Two basic ideas have been proposed to explain the origin of these objects: direct
stellar collisions \citep{hills76}, and mass transfer between or coalescence of the components 
of binary systems (\citealt{mcc64}; \citealt{chen08} and references therein)
 forming, respectively, collisional and evolutionary BSs. 
The role of each of these scenarios in producing the observed BSs populations is still
debated
(\citealt{fer12} and references therein; \citealt{chat13,per15}).
It is conceivable that different intracluster 
environments could be responsible for different origins of BSs: in less dense clusters they 
would preferentially form as evolutionary mergers of, or due to mass transfer in close 
binaries, while in high-density ones from collisions during dynamical encounters between 
binaries and/or single stars \citep{chat13,hyp13}.

BSs are not rare objects: every GC has been observed to host them, and each GC can contain 
from a few up to a few hundreds of them. The resulting 
straggler fractions range from a few times $10^{-5}$ to $10^{-4}$ \citep{pio04}, however their 
presence in a cluster is far from insignificant. They result from the fascinating interplay between 
stellar evolution and stellar dynamics which drives cluster evolution \citep{can15}, and as 
such they represent a crucial link between standard stellar evolution and internal dynamics 
of GCs \citep{fer15}. Thus, by studying them one can reconstruct the dynamical 
history of a cluster, and obtain important constraints on the role played by cluster dynamics 
on the evolution of cluster members. BSs statistics can also provide constrains on the initial 
distribution of binary systems (Hypki and Giersz 2013). All these factors explain why within 
the last twenty years BSs have evolved from a marginal oddity into an attractive subject of 
interest for at least three astronomical specialties: observers, cluster evolution modelers, 
and stellar/binary evolution modelers. 

Observational data on which our paper is based have been collected within the 
Cluster AgeS Experiment (CASE) project. The main aim of CASE is to determine 
the basic stellar parameters (masses, luminosities, and radii) of the components 
of GC binaries to a precision better than 1\% in order to measure ages and 
distances of their parent clusters, and to test stellar evolution models
\citep{jka05}. The best-suited objects for this type of survey are 
well-detached binaries so evolutionarily advanced that at least one component 
is about to leave (or has just left) the main sequence. A by-product of the 
systematic search for such systems is a wealth of cluster variables. Among the latter 
about 20 binary BSs have been found \citep{jka13a,jka13b,jka14,jka15,jka16} 
which strongly supports the mass transfer option. The aim of the present paper 
is to verify if a consistent evolutionary 
scenario can be obtained for such objects based on the available data. 

To the best of our knowledge, the four stragglers whose evolutionary history we attempt 
to reconstruct are the only ones in GCs whose 
parameters are accurately (or at least reasonably accurately) known. In Section 
\ref{sect:data} we discuss their properties together with observational limits on 
their chemical composition and age. Section \ref{sect:code} briefly introduces the 
evolutionary code employed to follow their evolution. Section \ref{sect:search} 
presents results of the search for their progenitors. The results are discussed in
Section \ref{sect:disc}, and the paper is summarized in Section \ref{sect:summ}. 

\section{Observational data}
\label{sect:data}

The subject of the present investigation is binary blue stragglers
V209/$\omega$ Cen, V60/M55, V228/47 Tuc and V8/NGC 6752 (hereafter:
V209, V60, V228 and V8, respectively).  The parameters of all the four
systems (orbital period $P$, semi-major axis of the orbit $a$,
component masses $M_1$ and $M_2$, component radii $R_1$ and $R_2$,
effective temperatures of the components $T_1$ and $T_2$) are
collected in Table~\ref{tab:data} together with ages and chemical
compositions or their parent clusters. We note that the quoted
temperature uncertainties include the spread of color-temperature
calibrations used, but accounting for discrepancies among 
  the various calibrations
  they might be increased to 250 -- 300 K for V60 (a synthetic
  calibration was employed; see \citealt{mnr13} for a justification)
  and V228 ($T_1$ was found iteratively, and its error was difficult
  to estimate, see \citealt{jka07b}).
 
\subsection{Binaries}

\subsubsection{V209}
V209 was discovered by \citet{jka96} during a search for variable stars in the 
field of $\omega$ Centauri. The light curve exhibits total eclipses and is nearly 
flat between the minima, indicating a detached system. A detailed account from 
photometric and spectroscopic observations can be found in \citet{jka07a}, who also 
derive absolute parameters of the components. On the CMD 
of the cluster (their Fig. 8), the primary component is located at the tip of the 
BSs region, while the much fainter ($\Delta V\approx1.5$ mag) secondary
approaches the hot subdwarf (sdB) domain. 

In the evolutionary scenario proposed by \citet{jka07a} V209 evolved through the common 
envelope phase when its original primary ascended the giant branch (RGB). The resulting angular 
momentum loss transformed it into a close binary with the orbital period of $\sim$1 d,
composed of a white dwarf and a main-sequence companion. The second episode of mass transfer 
and/or loss began once the companion entered the Hertzsprung gap. During that episode it lost 
most of its envelope, failing to ascend the RGB and ignite helium. Instead, it started to move 
nearly horizontally across the CMD toward the sdB domain. At the same time the primary accreted
enough mass to ignite hydrogen in a shell. Its envelope expanded, and the former white dwarf 
is currently seen as the more massive and more luminous component of V209 Cen. Its companion, 
which consists of a helium core surrounded by a thin hydrogen shell, is now seen as the less 
massive and hotter component of the binary. 
\citet{jka07a} admit that the chances of observing this evolutionary phase are rather low, 
as the primary should evolve very rapidly across the CMD, but they do not provide any 
alternative scenario.

\subsubsection{V60}
V60 was discovered by \citet{jka10}. The 
system is a semidetached Algol with eclipses of a very different depth ($\Delta 
V\approx1.5$ mag) whose period lengthens at a rate of $3.0\times10^{-9}$. On the 
CMD it occupies a position 
between the RGB and the extension of the main sequence 
\citep[Fig. 5 of][]{jka10}. \citet{mnr13} give a detailed account from photometric 
and spectroscopic observations, and derive absolute parameters of the components. 
They conclude that the present state of V60 is a result of rapid but 
conservative mass exchange which the binary is still undergoing.

\subsubsection{V228}
V228 was discovered by \citet{jka98}. The system is a semidetached Algol with 
partial eclipses of a significantly different depth ($\Delta V\approx0.3$ mag).
\citet{jka07b} give a detailed account from photometric and spectroscopic observations,
and derive absolute parameters of the components. On the CMD
of 47 Tuc (their Fig. 5), the primary component is located slightly blueward of 
the tip of the BSs region, while its companion resides close to the main sequence 
about $\sim$0.5 mag above the turnoff.
According to \citet{jka07b}, the secondary of V228 is burning hydrogen in a shell 
surrounding a degenerate helium core, and transfers mass to the primary on a nuclear
time scale. This means that the resulting rate of period lengthening should be very low,
and indeed: their observations with a time base of 11 yr failed to reveal any 
change of the orbital period.

\subsubsection{V8}
V8 was discovered by \citet{ibt99}. The system has a 
W UMa-type light curve with a total secondary eclipse. 
Spectroscopic data were not obtained, but the duration of totality together 
with the amplitude of light variation allowed to impose constraints on mass ratio $q$ 
and orbital inclination $i$. 
An account from observations is given by \citet{jka09}, who also perform a 
detailed analysis of the light curve. On the CMD (their Fig. 5) the primary of V8 
resides at the extension of the normal main sequence, while the secondary is 
located far to the blue of the main-sequence (the temperatures of both components 
are very similar). \citet{jka09} conclude that the present configuration of V8 
resulted from a substantial mass exchange between the components. Most likely, the 
present secondary lost nearly all matter from its original envelope, and is burning 
hydrogen in a shell.
 
We adopted the best-fitting model from among the radiative solutions of \citet{jka09}, and 
calculated absolute parameters of V8 based on apparent magnitude of the primary 
($V = 17.43$ mag) and distance modulus to NGC 6752 \citep[$\mu_V=13.13$ mag;][2010 edition]{har10}.
Their values are only approximate.

\subsection{Clusters}

Ages of the clusters together with their uncertainties are taken from \citet{marf09} and 
\citet{forb10}. Ages of 47 Tuc and NGC 6752 were more recently discussed by \citet{roed14}; 
however their values remained unchanged.
The Z-values in Table \ref{tab:data} were derived from $[\mathrm{Fe/H}]$ and $[\alpha/\mathrm{Fe}]$ 
indices quoted below using the XYZ calculator of 
G. Worthey.\footnote{http://astro.wsu.edu/models/calc/XYZ.html}

\subsubsection{$\omega$ Cen}
In $\omega$ Cen, which has been long known for its inhomogeneity, a broad range of Y and [Fe/H] 
is observed. According to a recent survey by \citet{frai15}, the cluster contains seven 
subpopulations with $0.25< \mathrm{Y}<0.4$, $-2.0<\mathrm{[Fe/H]}<-0.5$, $0.2<[\alpha/
\mathrm{Fe}]<0.6$ which differ from each other also in spatial distribution and kinematical 
properties.  As a result, one is free in choosing [Fe/H], 
$[\alpha/\mathrm{Fe}]$ and Y for evolutionary calculations of V209 (under the condition, of course, 
that they conform to the limits \citet{frai15} set for a given subpopulation).
In particular, the subpopulation OC3c on which we focus for reasons explained in Section 
\ref{sect:search} has $0.25< \mathrm{Y}<0.4$ and $-1.75<\mathrm{[Fe/H]}<-0.5$. 
We note that the spectra of \citet{jka07a} were not good enough to derive abundances, 
and they adopted [Fe/H] = -1.7 which represents the peak of the metallicity distribution in the cluster.

\subsubsection{M55}
The RGB of M55 is split on the ultraviolet CMD, suggesting 
the presence of two subpopulations \citep{pio15}. However, spectroscopic evidence speaks rather
for chemical homogeneity \citep{kay06,pan10}. We adopt [Fe/H] = -1.94 \citep[2010 edition]{har10},
$\mathrm{Y} = 0.25$, and a range 0.2 -- 0.4 for $[\alpha/\mathrm{Fe}]$.

\subsubsection{47 Tuc}
There is no doubt that 47 Tuc hosts more than one subpopulation of stars, although compared 
to $\omega$~Cen it may be regarded as nearly homogeneous. While all the subpopulations
have [Fe/H] = -0.72$\pm$0.01, at least one of them  may be enriched in He
up to $\mathrm{Y} = 0.28$ \citep{mar16}. We adopt $-0.73\le[\mathrm{Fe/H}]\le-0.71$, 
and $0.2 \le [\alpha/\mathrm{Fe}]\le 0.4$.

\subsubsection{NGC 6752}
In the case of NGC 6752 the problem of homogeneity is still discussed
\citep[see e.g.][and references~therein]{lap16}. The latter authors argue for the presence 
of three stellar subpopulations with Y = 0.24, 0.25 and 0.26. [Fe/H] values 
they quote range from -1.80 (based on FeI lines) to -1.50 (based on FeII lines). We adopt 
[Fe/H] = -1.54 \citet{har10}, $\mathrm{Y} = 0.25$, and - like for the other GCs - 
$0.2 \le [\alpha/\mathrm{Fe}]\le 0.4$. 
   \begin{table}
     \caption[]{Parameters derived from observations.}
     \label{tab:data}
     \begin{center}
         \begin{tabular}{l|c|c|c|c}
            \hline
Object  &  V209  &  V60  &  V228 &  V8  \\
  \hline                                                              
$P$(days)  &  0.834419  &  1.183021  &  1.150686  &  0.314912  \\       
$a/R_\odot$  &  3.838(55)  &  5.487(47)  &  5.529(24)  &  1.92  \\
$M_1/M_\odot$  &  0.945(43)  &  1.259(25)  &  1.512(22)  &  0.84  \\
$M_2/M_\odot$  &  0.144(08)  &  0.327(17)  &  0.200(07)  &  0.12  \\
$R_1/R_\odot$  &  0.983(15)  &  1.102(21)  &  1.357(19)  &  0.88  \\
$R_2/R_\odot$  &  0.425(08)  &  1.480(11)  &  1.238(13)  &  0.45  \\
$T_1$(K)   &  9370(300)  &  8160(140)  &  8075(130)  &  6990  \\
$T_2$(K)   &  10870(320) &  5400(160)  &  5810(150)  &  7100  \\
$\alpha_{2000}$ & 201.57200     &   294.99046    &   6.53692      & 287.79417  \\
$\delta_{2000}$& -47.38800     &   -30.96288    &  −72.11706     & -59.98151  \\
  \hline                                                              
Cluster  &  $\omega$ Cen  &  M55  &  47 Tuc  &  NGC 6752  \\
  \hline                                                              
age(Gyr)  &  11.5$\pm$0.9  &  12.3$\pm$0.5  &  13.1$\pm$0.9  &  11.8$\pm$0.6  \\
Y  &  0.25$\div$0.40  &  0.25  &  0.25$\div$0.28  &  0.25  \\
Z$/10^{-4}$  &  1$^a\div$110  &  1$\div$3  &  24$\div$52  &  4$\div$8  \\
            \hline
         \end{tabular} 
\end{center}
$^a$ For the OC3c subpopulation the lower limit is 3.
   \end{table}

\section{Evolutionary model of a cool close binary}
\label{sect:code}

The following search for progenitors of the investigated binaries is based on a model
developed by one of us \citep{ste06a, gste08, ste09, stekir15}. It
describes the evolution of a cool close binary from the zero-age main sequence
(ZAMS)  until a stage preceding the merger of the components.

The basic equations of the model are Kepler's Third Law, expression
for angular momentum (AM) conservation, and approximate expressions
for inner Roche-lobe sizes $r_1$ and $r_2$ \citep{eggl83}:

\begin{equation}
P = 0.1159a^{3/2}M^{-1/2}\,,
\label{kepler}
\end{equation}

\begin{equation}
H_{\rm{tot}} = H_{\rm{spin}} + H_{\rm{orb}}\,,
\label{totam}
\end{equation}

where 

\begin{equation}
H_{\rm{spin}} = 7.13\times10^{50}(k^2_1M_1R^2_1 +k^2_2M_2R^2_2)P^{-1}\,,
\label{spin}
\end{equation}

and

\begin{equation}
H_{\rm{orb}} = 1.24\times 10^{52}M^{5/3}P^{1/3}q(1+q)^{-2}\,,
\label{orbam}
\end{equation}

\begin{equation}
\frac{r_1}{a} = \frac{0.49q^{2/3}}{0.6q^{2/3}+\ln(1 + q^{1/3})}\,,
\end{equation}

\begin{equation}
\frac{r_2}{a} = \frac{0.49q^{-2/3}}{0.6q^{-2/3}+\ln(1 +
q^{-1/3})}\,.
\end{equation}
Here $M = M_1 + M_2$ is the total mass of the binary, $H_{\rm{tot}}$, $H_{\rm{spin}}$ 
and $H_{\rm{orb}}$ are total, rotational and orbital AM, $k_1^2$ and 
$k_2^2$ are the (nondimensional) gyration radii of the components, $r_1$ 
and $r_2$ are the sizes of the inner Roche lobes, and $q = M_1/M_2$ is 
the mass ratio. Mass, radius, and semi-major axis are given in solar units, period 
in days, and AM in cgs units. 

We assume that orbital and rotational 
motions are fully synchronized throughout the whole evolution, i.\thinspace e. that 
the rotational period of each component is always equal to the orbital period
$P$. This assumption is very well founded for close binaries based both on
  observations and theoretical considerations: systems with $P<~10$ are fully 
  synchronized \citep {abt02}, and synchronization time for periods shorter 
  than 2-3 d is of the order of $10^4 - 10^5$ years \citep {zahn89}.

We consider the evolution of an isolated binary, not interacting
dynamically with any other object. So, stellar winds from the two
components and the mass transfer between them, are the
dominant mechanisms of the orbit evolution. The winds carry
away mass and AM according to the formulae

\begin{equation}
\dot M_{1,2} = -10^{-11}R_{1,2}^2\,,
\label{massloss}
\end{equation}

\begin{equation}
\frac{{\rm d}H_{\rm{tot}}}{{\rm d}t} = -4.9\times 10^{41}
(R_1^2M_1 +R_2^2M_2)/P\,.
\label{amloss}
\end{equation}

Here $\dot M$ is in solar masses per year and ${{\rm
    d}H_{\rm{tot}}}/{{\rm d}t}$ is in gcm$^2$s$^{-1}$ per year. The
formulae are calibrated by the observational data of the rotation of
single, magnetically active stars of different age, and empirically
determined mass-loss rates of single, solar type stars. Both formulae
apply in a limiting case of a rapidly rotating star in the saturated
regime. Note that they do not contain any free adjustable
parameters. The constant in Eq.~(\ref{massloss}) is uncertain within a
factor of 2 and that in Eq.~(\ref{amloss}) is uncertain to $\pm$ 30\%
\citep{ste06b,wood02}. The model ignores any interaction between winds
from the two components.

The evolutionary calculations are divided into three phases: from ZAMS
to the Roche lobe overflow (RLOF) by the initially more massive component
(henceforth donor), rapid, conservative mass exchange from the more to
less massive component (henceforth accretor), and the last phase of a
slow mass exchange resulting from nuclear evolution of the donor.

The present computations follow those described in
\citet{stekir15}, so the reader is referred to that paper for
additional details, assumptions etc. However, a couple of
modifications were introduced  to increase accuracy and computational efficiency 
of the code. In particular, upon approaching the final evolutionary state the time step
was substantially shortened to avoid nonphysical period oscillations observed in some 
cases by \citet{stekir15}. Instead of Padova models \citep{gir00}, the newer PARSEC 
models\footnote{http://stev.oapd.inaf.it/cgi-bin/cmd} \citep{bre12} were used, which 
include up-to-date physics and cover a broader range of metallicities. Finally, for 
modeling of V209 and V228 we calculated our own evolutionary tracks using a procedure 
outlined in Section \ref{sect:V209s}.

\section{The search for progenitors of the investigated binaries}
\label{sect:search}

\begin{figure}
\includegraphics[height=\hsize]{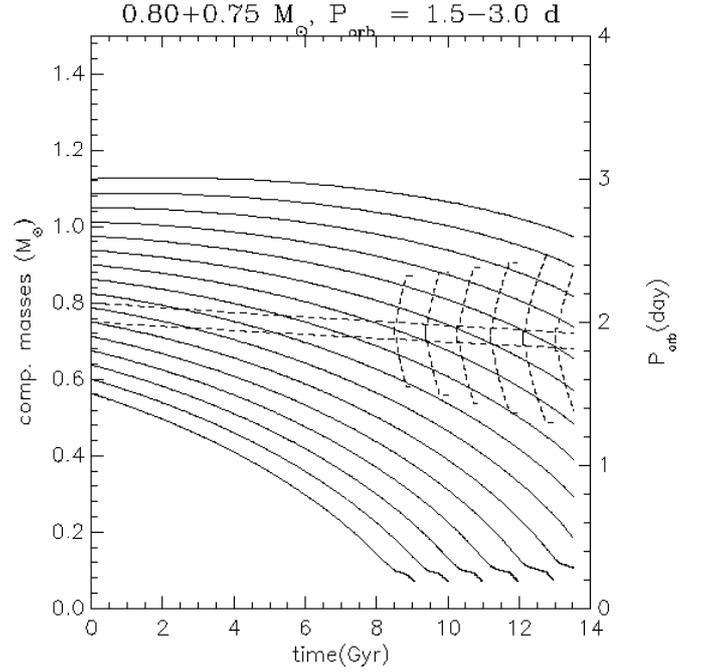}
\caption{Time evolution of Set 1 models with initial component
  masses of 0.8 and 0.75 $M_{\sun}$ (broken lines, left axis) and 
  initial periods of 1.5-3.0 d (solid lines, right axis). Only the 
  six shortest-period binaries reach the RLOF phase within the 
  age of the Universe, enabling the mass exchange between their
  components. All other binaries stay detached.}
\label{m0.80}
\end{figure}

\begin{figure}
\includegraphics[height=\hsize]{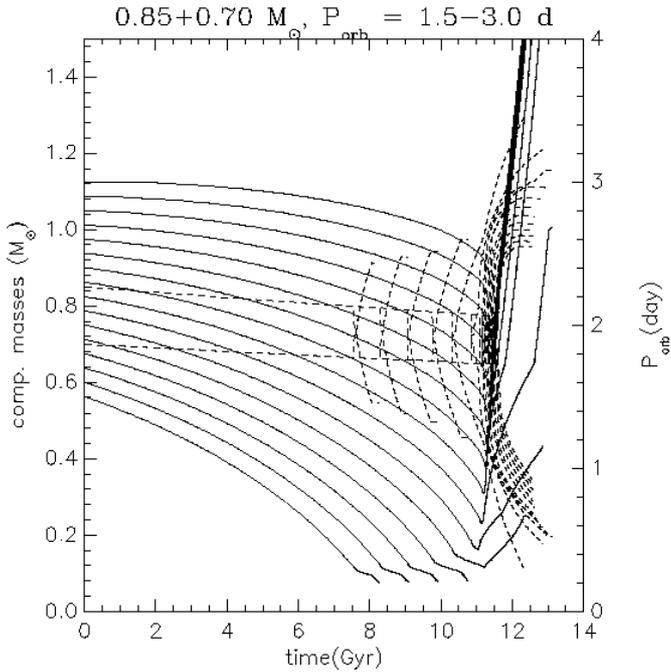}
\caption{Same as in Fig.~\ref{m0.80} but for Set 2 models with initial component
  masses of 0.85 and 0.70 $M_{\sun}$. Only the four shortest-period binaries
  enter the RLOF phase when the primary component is still on the main sequence. 
  For all other binaries RLOF occurs when the primary is past the main sequence
  i.\thinspace e. at an age of $\sim$11 Gyr.}
\label{m0.85}
\end{figure}

\begin{figure}
\includegraphics[height=\hsize]{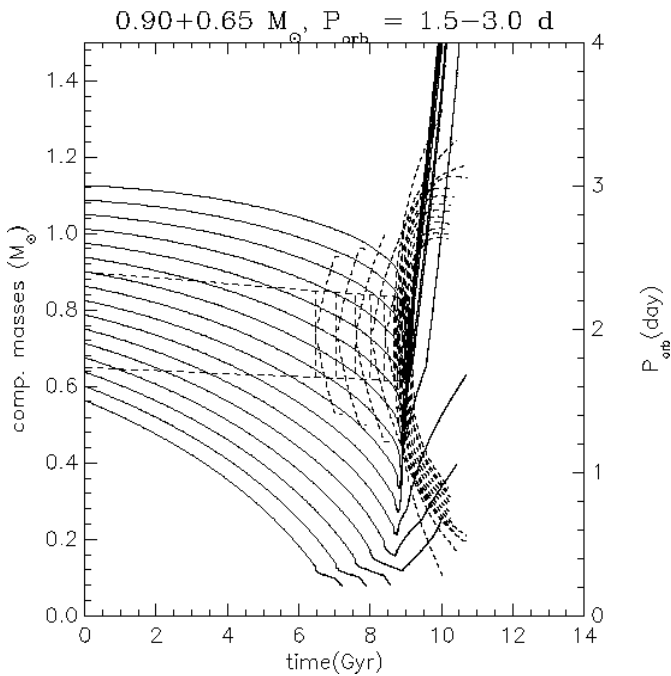}
\caption{Same as in Fig.~\ref{m0.80} but for Set 3 models with initial component masses
  equal to 0.90 and 0.65 $M_{\sun}$. Only the three shortest-period binaries
  enter the RLOF phase when the primary component is still on the main sequence.        
  For all other binaries RLOF occurs when the primary is past MS i.\thinspace e. 
  at an age of $\sim$9 Gyr.}

\label{m0.90}
\end{figure}

Equations from the previous Section do not contain any free adjustable
parameters, so after specifying helium (Y) and metal (Z) content, the
initial masses of both components and the initial orbital period, the
evolution of the binary is fully determined.  Our task is to
examine the initial parameters space until a model binary fitting
observations is found. Note that the result of the search is not
unique if the physical parameters of the observed binary,
i.\thinspace e. chemical composition, orbital period, masses, sizes
and luminosities of the components, are only used for matching
\citep{ste11}. Several model binaries with different values of the
initial parameters can reproduce the observed values but in each case
a different time is needed to reach the agreement with the
observations.  

To demonstrate the nonuniqueness of the solution in a broader context 
  we computed three sets of models with initial mass ratios $q_\mathrm{init}$ 
  = 0.94, 0.82 and 0.72; all of them with initial total mass $M_\mathrm{init} 
   = 1.55 M_{\sun}$, and low metallicity Z = 0.001 characteristic of globular clusters 
  ($M_\mathrm{init}$ was kept fixed assuming that a progenitor of an exemplary binary 
  with the final total mass close to 1.4 $M_{\sun}$ was searched for). 
  Each set contained 16 binaries with initial orbital periods of between 1.5 d and 3 d. 
  In Set 1, the initial component masses were equal to (0.8,0.75) $M_{\sun}$; in 
  Sets 2 and 3 - to (0.85,0.7) and (0.9,0.65) $M_{\sun}$, respectively. Note that 
  accounting for mass loss due to the wind, the MS-lifetime of 
  a 0.8 $M_{\sun}$ star exceeds the age of the Universe, while that of 
  0.85 $M_{\sun}$ and 0.9 $M_{\sun}$ is shorter than the age of most GCs. 

  As it is seen in Fig.~\ref{m0.80}, the 0.8
  $M_{\sun}$ star fills its inner Roche lobe only in binaries with initial
  orbital periods shorter than 2.1 d. The orbits of these binaries are
  then so compact that soon afterwards both the components overfill the outer 
  critical Roche surface and merge together.
  For periods equal to, or longer than 2.1 d RLOF never
  occurs and the binaries stay detached. Set 1 models can only reproduce 
  observed contact or near-contact binaries with parameters from a restricted 
  range $M_1 \la 0.9 M_{\sun}, M_2 \ga 0.55 M_{\sun}$ and $P \la 0.4$ d. 

  The situation changes a lot when we consider Set 2 binaries with $M_1 = 0.85 M_{\sun}$
  (Fig. \ref{m0.85}). An $M_1 = 0.85 M_{\sun}$ primary 
  fills the Roche lobe during the MS evolutionary phase only in binaries with the 
  four shortest periods. For all other periods it leaves the main sequence while 
  being detached from its Roche lobe, but the ensuing expansion quickly leads to 
  the RLOF. A rapid mass exchange occurs, after which the orbital period increases 
  and the mass ratio decreases. Set 2 models can reproduce observed binaries with 
  parameters from a broad range $M_1 \la 1.25 M_{\sun}, M_2 \ga 0.2 M_{\sun}$,
  and $P$ from a fraction of a day up to a few days. Their final ages are concentrated 
  around 11 Gyr.
  Fig.~\ref{m0.90} shows the evolution of Set 3 models with $M_1 = 0.9 M_{\sun}$.
  It is very similar Fig.~\ref{m0.85}, except that the final ages are concentrated 
  around 9 Gyr.

\begin{figure}
\includegraphics[height=\hsize]{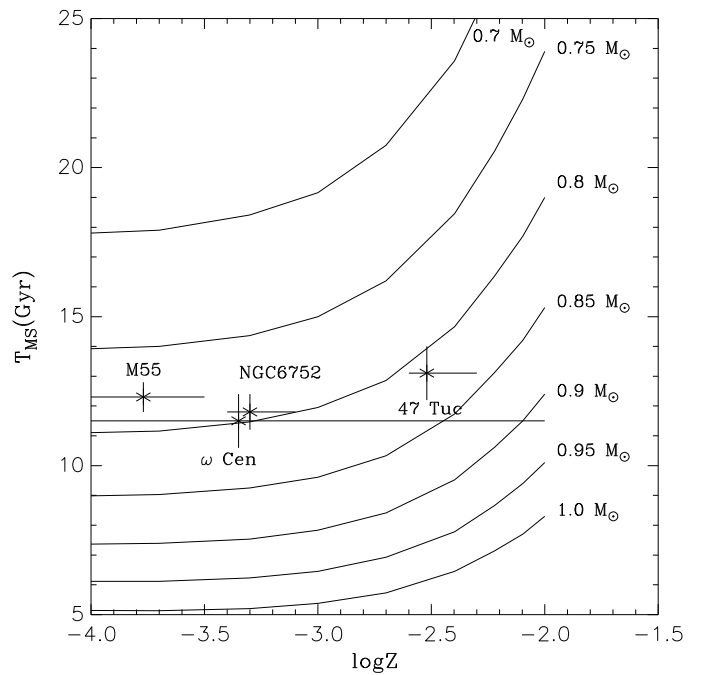}
\caption{The relation between metallicity and life-time on the MS for stars
  with different masses, based on PARSEC models. Indicated are the four 
  clusters discussed in the present paper. Vertical and horizontal line 
  segments give for each cluster uncertainties of both coordinates 
  (see Table~\ref{tab:data}).}
\label{agevsz}
\end{figure}

 As we see, the final age of the model crucially depends on the initial mass 
 of the primary, $M_1$. Changing Z has a similar, but weaker effect 
 as changing $M_1$ (see Fig.~\ref{agevsz}). Knowledge of age and
 metallicity of the observed binary lifts the indeterminacy of the initial
 parameters. Fortunately, for GC members these parameters are known with a
 reasonable accuracy, even in clusters containing multiple subpopulations
 \citep{roed14}.

Figs.~\ref{m0.80}-\ref{m0.90} may be regarded as an illustration of a
coarse search of the parameter space. To reproduce the observed binary
parameters more accurately, all the three initial parameters have to
be varied within a narrow range around the values resulting from the
coarse search. This also includes the total initial mass because
  different component masses result in somewhat different mass
  losses. Examples of such fine search in period are shown in
Figs. \ref{v209}-\ref{v8} related to the four actual binaries
discussed below.

\begin{table*}
\caption{Data on observed and model binaries.}
\label{data}
\centering 
\begin{tabular}{l|ccc|ccc|ccc|ccc}
\hline
\hline
& \multicolumn{3}{|c|}{V209} &
\multicolumn{3}{|c|}{V60} &
\multicolumn{3}{|c|}{V228}&
\multicolumn{3}{|c}{V8} \\
\hline
Parameter & obs. & mod.(1) & $\Delta$(o-m)$^*$ & obs. & mod.(2) &
$\Delta$(o-m)$^*$ & obs. & mod.(3) &  $\Delta$(o-m)$^*$ & obs. &
mod.(4) & $\Delta$(o-m)$^*$\\
\hline
age(Gyr)       &    11.5 &  11.34 & 1.4\% &  12.3 &   10.13 & 17.6\% &
13.1 &  10.39 & 20.6\% &  11.8 &  11.58 & 1.9\%\\ 
$P$(days)      &   0.834 &  0.834 & 0.0 &  1.183 &  1.183 & 0.0 &
1.151 &  1.151 & 0.0 &  0.315 &  0.315 & 0.0\\
$M_1/M_{\sun}$    &    0.945 &  0.959 & -1.5\% &  1.259 &  1.251 &
0.6\% &  1.512 &
1.509 & 0.2\% &  0.84 &   0.834 & 0.7\%\\
$M_2/M_{\sun}$   &    0.144 &  0.149 & -3.3\% &  0.327 &  0.326 &
0.3\% &  0.200 &
0.197 & 1.5\% & 0.12 &   0.125& -4.2\%\\
$R_1/R_{\sun}$    &     0.983 &  0.998 & -1.6\% &  1.102 &  0.994 &
9.8\% &  1.357 &
1.434 & -5.7\% & 0.88 &   0.74& 15.9\%\\
$R_2/R_{\sun}$    &     0.425 &   -  & - &    1.480 &  1.497 & -1.1\%
&  1.238 &
1.246 & -0.6\% & 0.45 &   0.45& 0.0\\
$\log(L_1/L_{\sun})$ &    0.82 &   0.84 & -0.02 &   0.68 &   0.77 &
-0.09 &   0.85 &   1.04 & -0.19 &   0.21 &   -0.10 & 0.31\\
$\log(L_2/L_{\sun})$ &    0.35 &    -   & - &   0.22 &   0.31 & -0.09
&   0.20 &   0.15 & 0.05 &   -0.37 &   - & -\\
$\log{T_1}$(K)   &  3.97 &   3.97 & 0.0 &   3.91 &   3.96 & -0.05 &
3.91 &   3.94 & -0.03 &   3.84 &   3.80& 0.04\\
$\log{T_2}$(K)   &  4.04 &    -  & - &   3.73 &  3.75 & -0.02 &
3.76 &   3.75 & 0.01 &   3.85 &    - & -\\

\hline
\end{tabular}
\begin{flushleft}
{$^*$ In the last four rows differences of logarithms are given\\
mod.(1): computed model with Y = 0.40 and Z = 0.0004. Initial
  parameters: ($M_1,M_2)_\mathrm{init} = (0.65, 0.55) M_{\sun}$ and
  $P_{\mathrm{init}}$ = 1.66 d\\ 
mod.(2): PARSEC model with Y = 0.25 and Z = 0.0002. Initial
  parameters: ($M_1,M_2)_\mathrm{init} = (0.87, 0.86) M_{\sun}$ and
  $P_{\mathrm{init}}$ = 2.045 d\\ 
mod.(3): computed model with Y = 0.27 and Z = 0.003. Initial
  parameters: ($M_1,M_2)_\mathrm{init} = (0.95, 0.94) M_{\sun}$ and
  $P_{\mathrm{init}}$ = 2.129 d\\ 
mod.(4): PARSEC model with Y = 0.25 and Z = 0.0005. Initial
  parameters: ($M_1,M_2)_\mathrm{init} = (0.85, 0.19) M_{\sun}$ and
  $P_{\mathrm{init}}$ = 3.715 d}
\end{flushleft} 
\end{table*}

To select the best-fitting models, we applied the following procedure:
first, we adopted the Y and Z parameters in agreement with the
observations of the parent clusters. Then, we evolved the model
  binary until the orbital period was to the third decimal place equal
  to that listed in Table \ref{tab:data}, and, simultaneously, the
  observed component masses were reproduced within the observed
  uncertainties. This accuracy of period fitting proved high enough
  for the resulting uncertainty of progenitor parameters being
  negligible compared to the overall uncertainty dominated by sources
  discussed in the following Sections. Finally, of all models meeting
  these criteria the one with the age closest to the cluster age was
  selected. Radii, effective temperatures and luminosities of the
components resulted automatically from the evolutionary sequences.
The exception, discussed in detail in Section \ref{sect:V209s},
  were extremely low-mass stars $(M< 0.2M_{\sun})$ with helium cores
  and extended hydrogen envelopes which we could not model reliably
  enough.

The basic data for the best-fitting models are given in
Table~\ref{data}. For each star the left column contains age of the
parent cluster and the observed values of stellar parameters. The age
of the model is given in the first row of the middle column and then
all other parameters, as obtained from the model. For the secondary components
of V209 and V8 no data on temperature and luminosity are given because 
we were not able to model extremely low mass stars with a substantial
helium core. There exists a well known
relation connecting the mass of the helium core with the stellar
radius and luminosity but it applies only to red giants with a
sufficiently high total and core mass \citep{iben93,eggl06}. Stars
less massive than about 1 $M_{\sun}$ with the
helium core mass lower than about 0.2 $M_{\sun}$ obey individual core
mass-luminosity relation \citep[see Fig. 2.9 in][]{eggl06} as well as
core mass-radius relation \citep{mus96} depending on the total stellar
mass. The right column gives for each star relative differences 
$\Delta$(obs.-mod.) expressed in per cent. 


The ages of V209 and V8 are close to the adopted ages of the hosting
clusters $\omega$ Cen and NGC 6752, but a serious mismatch occurs in
the remaining two cases, in which the best-fitting models are
significantly younger than the corresponding clusters. The discrepancy
is of fundamental nature, as Fig.\ref{agevsz} shows. Here we plotted
the MS life time of stars with different masses and metallicity. The
lines of constant mass are shown with solid lines and labels. We also
plotted the positions of all four clusters according to their age and
metallicity. The positions give the maximum mass of a star which just
reached TAMS in each cluster. This mass is close to, but not identical
with the turn-off mass. Specifically, the turn-off mass, defined
  as a hottest MS star on the isochrone corresponding to a given
  cluster, is lower by 0.01-0.02 $M_{\sun}$ than the plotted maximum
  mass. Doubling this mass results in the maximum total mass of a
close binary which can still be found in the blue straggler region of
the cluster (assuming a strictly conservative evolution, i.e. no mass loss
from the system and no close interactions with other members of the cluster). The
read-off values of the maximum mass are: $0.78 \pm 0.01 M_{\sun}$ for
M55, $ 0.8-0.9 M_{\sun}$ for $\omega$ Cen, $0.80 \pm 0.01 M_{\sun}$
for NGC 6752 and $ 0.81 \pm 0.01 M_{\sun}$ for 47 Tuc. While the
observed total masses of the analyzed binaries in $\omega$ Cen and NGC
6752 are lower than these limits, they are higher in two other
clusters: 1.586 versus 1.56 $M_{\sun}$ for M55 and 1.712 versus 1.62
$M{\sun}$ for 47 Tuc. Moreover, the evolutionary model includes the
mass loss of several percent during the binary evolution (mostly
during the longest first phase), so the initial component masses must
be even higher than the measured ones (see Table~\ref{data}). In
effect, no model binary can be found that reproduces the observed
total mass and age of V60 and V228 (assuming, of course, that their
ages are equal to the ages of their parent clusters).

\subsection{V209}
\label{sect:V209s}

\begin{figure}
\includegraphics[height=\hsize]{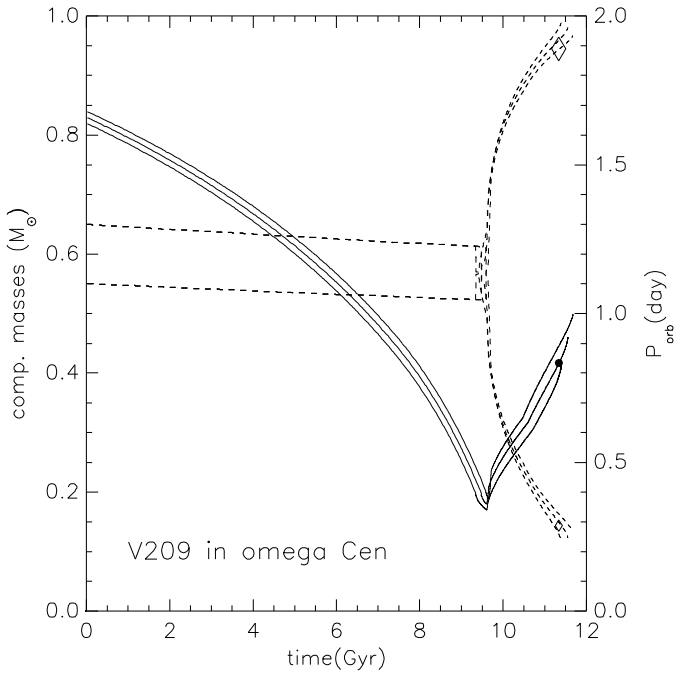}
\caption{Time variations of orbital period (solid lines, right
  axis) and component masses (broken lines, left axis)
  of the best fitting model of V209 (see Table~\ref{data} 
  for its initial parameters). Also shown are two models with 
  the same initial component masses, and periods differing by 
  $\pm$0.02 d from the best-fitting one. The observed parameters 
  of V209 are indicated by a filled circle (period) and diamonds (masses). 
  Vertical sizes of the diamonds correspond to observational errors. }
\label{v209}
\end{figure}

$\omega$ Cen is the most massive but also the most unusual GC
in our Galaxy. It shows a very complex structure with a very
broad MS, suggestive of a combination of a few independent sequences,
multiple horizontal, subgiant and red giant branches, and a large
spread in heavy element content \citep[][and references
therein]{nor97, john10, pio05, joo13, frai15}. As a result, several
(up to seven) different subpopulations have been identified with
different chemical compositions and ages. In particular, an enhanced
helium abundance was suggested for two subpopulations \citep{joo13, frai15}. 

To find a progenitor of V209, we first used PARSEC models in a search
for desirable initial binary parameters. It quickly turned out that
at the helium abundance Y = 0.25 -- 0.27 the primary of V209 is far too hot 
for its mass, independently of metal content and age varying within the
  observed values. It is known, however, that evolutionary tracks
    of the helium enriched models are shifted towards higher
    temperatures (and higher luminosities) in the CMD, compared to normal helium abundance
    models \citep{glebb10}. Based on the suggestion
that a significant subpopulation of cluster members may have helium
content Y $\approx$ 0.40 \citep{joo13, frai15} we 
generated our own grid of evolutionary tracks for helium-enriched stars.

The tracks were computed using the Warsaw-New
Jersey evolution code which is a modern version of B.
Paczy\'nski's code \citep{bep69,bep70}. The computations were
performed starting from chemically uniform models on the ZAMS. 
The input parameters for evolutionary model
sequences are total mass, $M$, initial value for helium abundance, $Y_0$, and
heavy element abundance, $Z$. We computed two series of models with masses
from 0.50 to 1.60~$M_\odot$, with a step size of 0.05~$M_\odot$. In one series
of the computations we assumed $Y_0 = 0.40, Z = 0.0004$, in the second one -
$Y_0 = 0.27, Z = 0.003$. For chemical elements heavier than helium, we adopted
the solar element mixture of \citet{asp09}. For the opacities, we used the OP data
\citep{se96,se05} supplemented at $\log T < 3.95$ with the low-temperature
data of \citet{fer05}. In all computations the OPAL equation of
state was used \citep[version OPAL~EOS~2005]{rog96}.
The nuclear reaction rates are the same as used by \citet{bah95}.
In the stellar envelope, the standard mixing-length theory of convection
with the mixing-length parameter $\alpha=1.6$ was used.
The choice of the mixing-length parameter is important in our models
because they are sufficiently cold to have an effective energy transfer
by convection in the stellar envelope, namely in the regions of partial ionization of
hydrogen and helium, and therefore will have influence on the structure of the models and their
position on the HR diagram. We used value of $\alpha$ to be close to estimations for the Sun
between 1.7 and 2.0 \citep{mag15} as obtained from 3D computations
and calibration to the solar radius.

Based on those models we were able to find a progenitor of V209 (see
Table~\ref{data}). A binary with the initial orbital period of 1.66 d
and the component masses of 0.65 and 0.55 $M_{\sun}$ evolves after
11.34 Gyr (which is almost equal to the adopted age of $\omega$ Cen)
into a binary with parameters close to V209, see Fig.\ref{v209}.
  Tracks for additional two models are also plotted to illustrate the
  accuracy of the fine parameter search, in this case in period.
 The models have the same
  initial component masses as the best-fitting one, but their initial
  periods differ by $\pm$0.02 d from the best-fitting one. The
  obs.-mod.  differences are significantly larger for them: $\Delta P$
  = 0.8\% and 6.8\%, $\Delta M_1$ = 3.7\% and 0.1\% and
  $\Delta M_2$ = 8\% and 13\% for the shorter and longer
  period model, respectively. A change of either component mass by e.\,g.
  ~1\% results in a similar worsening of the fit. We conclude that
  fitting the model to the observations with the assumed accuracy
  requires a really fine tuning of the initial parameters.

 We note that in our model, unlike in the scenario proposed
by \citet{jka07a}, there is no need for the binary to experience two CE
phases: just one mass transfer phase is sufficient to reproduce all 
the observed parameters of the primary (accretor) within the observational
uncertainty. Unfortunately, in case of the secondary (donor) we could
only compute its mass and the mass of its helium core. We followed the
evolution of the donor until its mass approached 0.2 $M_{\sun}$,
assuming that it still fills its Roche lobe. The helium core then reached
a mass of about 0.13 $M_{\sun}$. As the mass of the star decreases beyond
0.2 $M_{\sun}$ its radius shrinks rapidly, the star detaches from the
Roche lobe and loses mass via stellar wind and during the
thermonuclear flashes occurring in the hydrogen burning shell
\citep{sarna96, mus96, dri98, alth13}. Below about 0.18 $M_{\sun}$
flashes disappear but the radius (and mass) continues to decrease as
the star moves to the high temperature region in the HR diagram. These
phases are not described by our model because of lack of sufficiently
dense grid of evolutionary models of such stars. 
Nevertheless, the precise fit of
our model to the observed parameters of the primary component of V209,
in particular an unusually high effective temperature for its mass,
may be regarded as an independent and strong evidence for the existence 
of a stellar subpopulation in $\omega$ Cen with the high helium content 
of Y = 0.40.

\begin{figure}
\includegraphics[height=\hsize]{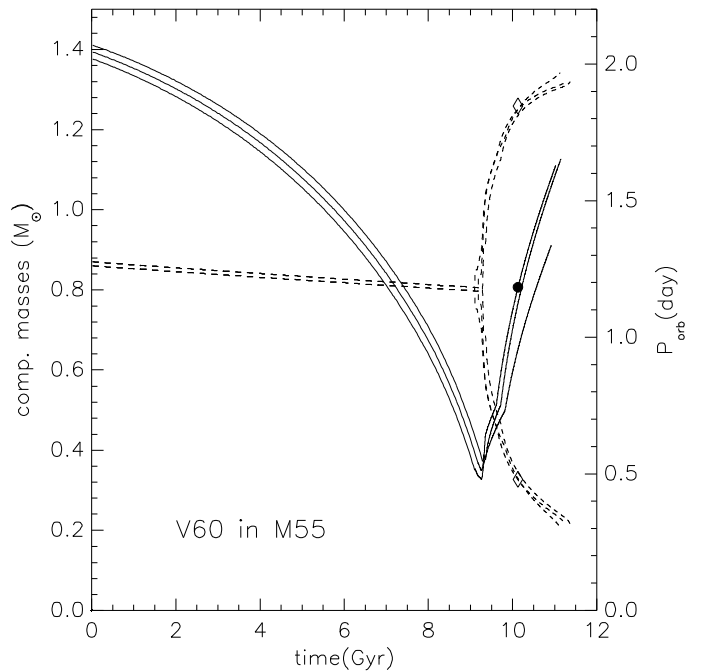}
\caption{The same as in Fig.\ref{v209} but for V60 in M55. The two additional
  models have initial periods differing by $\pm$0.025 d from the best-fitting one.}
\label{v60}
\end{figure}

\subsection{V60}

This binary was already used by \citet{stekir15} to demonstrate the potential of 
the method outlined in Section~\ref{sect:code} in identifying progenitors of close
binaries in GCs. Among models calculated in that paper, the one with $P_\mathrm{init}=1.9$~d 
and component masses $(M_1,M_2)$ = (0.89, 0.80)~$M_{\sun}$ reproduced the observed parameters 
of V60 fairly well. However, it could not be regarded as a solution to the present problem, 
as its metallicity did not conform to the limits given in Table~\ref{data}, and, moreover, the 
whole BS population of \citet{stekir15} was generated without paying attention to the age of
the models. Thus, a standard search of the parameter space was necessary. Upon finishing it, 
we found that the model best matching the present values $P, M_1$ and $M_2$ was over 2~Gyr 
younger than M55. As we argued at the beginning of Section \ref{sect:search}, this discrepancy
is a direct consequence of the doubled turnoff mass of M55 being smaller than the total mass of V60. 
The results of the present fit are shown in Fig.~\ref{v60}.

The CMD of the cluster suggests a presence of subpopulations differing in age and/or chemical 
composition \citep{pio15}. An age discrepancy of 2 Gyr between them seems rather unlikely, but 
it can not be entirely excluded. On the other hand, the ages of M55 and V60 could be reconciled 
if the metallicity of the straggler were significantly 
higher than the upper limit given in Table \ref{sect:data}. Fig.\ref{agevsz} shows that the present 
total mass of V60 can be reached for Z $\approx$ 0.005 i.\, e. about 15 times higher than obtained 
for the cluster (note that M55 is the most metal-poor of all GCs discussed in this paper). 
If V60 belonged to a subpopulation younger by 1 Gyr than the cluster, the value of Z necessary for
the turnoff mass to be equal to the total mass of the binary would decrease to 0.003 - a value still 
much larger than observed. Clearly, more data and better models are needed to resolve the discrepancy.

\subsection{V228}

\begin{figure}
\includegraphics[height=\hsize]{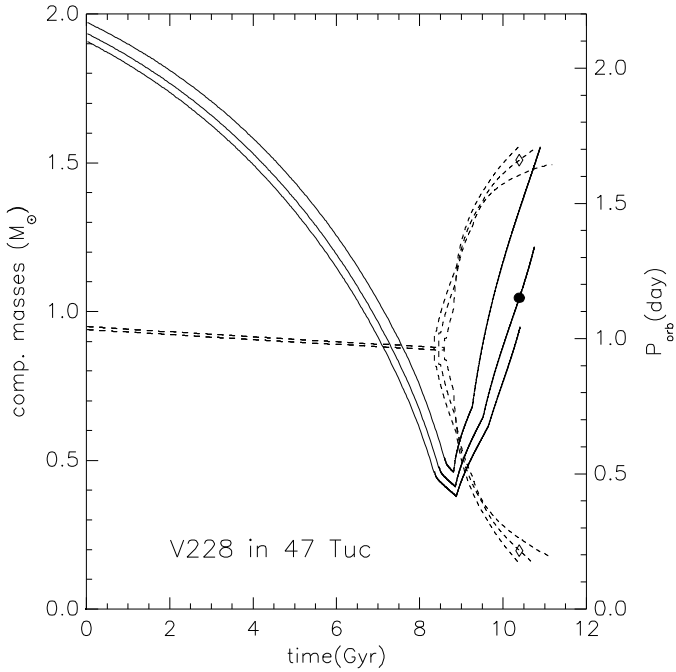}
\caption{The same as in Fig.\ref{v209} but for V228 in 47 Tuc. The two additional
  models have initial periods differing by -0.02 and 0.04 d from the best-fitting one.}
\label{v228}
\end{figure}

This is another variable star for which a substantial discrepancy
occurs between the adopted age of the cluster and the age of the best
fitting model (Table~\ref{data} and Fig.\ref{v228}). The PARSEC models with the
metallicity of 0.003 are not available, so we decided to compute a
special set of models with this value of Z and a helium
content Y=0.27. The best fitting model has the initial component masses
equal to 0.95 and 0.94 $M_{\sun}$, orbiting each other with the period
of 2.129 d.

A binary with such a high total mass, and metallicity
characteristic of 47 Tuc, should be past the red giant phase long
before the age of the cluster. The discrepancy between the cluster and
model age is here even larger than for M55 (2.7 vs. 2.1 Gyr). Similarly to other
clusters, 47 Tuc also shows
multiple subpopulations \citep{joo13, pio15} but so far no data on the
age spread are available. Alternatively, the age of the model binary can be
brought to agreement with the cluster age if its components have
metallicity 3-4 times higher than adopted for the cluster.

The secondary of V228 reached a mass of 0.2 $M_{\sun}$ but it still
fills its Roche lobe and has characteristics of a subgiant 
with a helium core of 0.12-0.13 $M_{\sun}$. The model reproduces the 
observed parameters quite well. We expect the star to detach soon from its Roche lobe
and to move to the left in the HR diagram.  The agreement of the model
with the primary of V228 is somewhat poorer. In particular, the radius
of the accretor is significantly larger than observed, and also the
temperature is significantly higher. As a result, the luminosity of
the model primary is too high by 0.19 dex, as compared to the observations. The
difference suggests that the real primary is {\em less}
evolutionarily advanced than the model.

\begin{figure}
\includegraphics[height=\hsize]{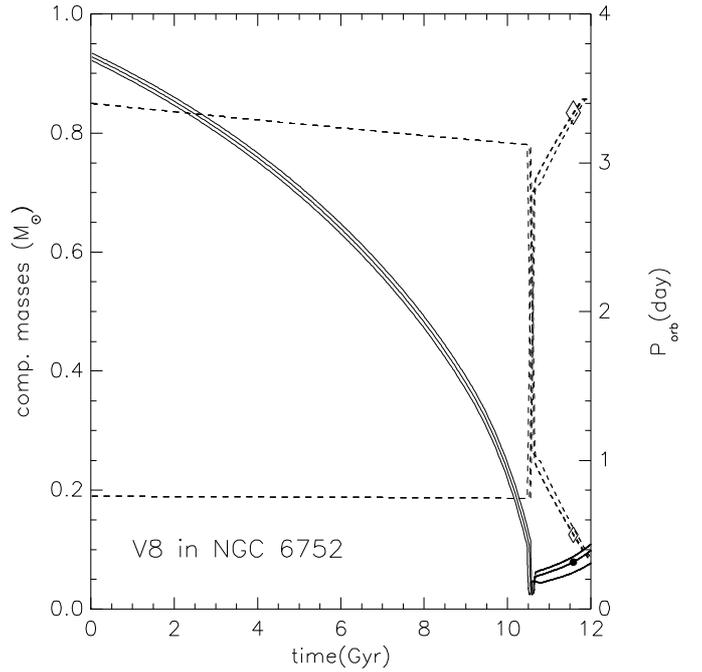}
\caption{The same as in Fig.\ref{v209} but for V8 in NGC 6752. The two additional 
  models have initial periods differing by $\pm$0.025 d from the best-fitting one.}
\label{v8}
\end{figure}

\subsection{V8}

This is the most unusual binary system of all discussed in the present
paper. It is a near contact binary with an extremely low mass
secondary of 0.12 $M_{\sun}$ possessing the radius of 0.45 $R_{\sun}$
and filling its Roche lobe. A single star with this mass and a
significant helium core has a much lower radius as the calculations by
\citet{mus96} show. A similarly low radius results from the paper by
\citet{pacz07} who computed radii of several stars with masses between
0.18-0.14 $M_{\sun}$ and different helium cores. The maximum stellar
radius decreases from about 1.3 $R_{\sun}$ for a 0.18 $M_{\sun}$ star
down to about 0.3 $R_{\sun}$ for a 0.14 $M_{\sun}$ star. An
extrapolation of this trend to 0.12 $M_{\sun}$ gives a maximum radius
of about 0.15 $R_{\sun}$, in a good agreement with the paper by
\citet{mus96}. The observed large value of the radius must result from
the interaction with the other component. V8 may remind in this
respect AW~UMa \citep{ruc15} where the 0.13 $M_{\sun}$ secondary
possesses a massive helium core and is submerged in the matter flowing
from the 1.3 $M_{\sun}$ primary component. AW~UMa looks apparently as
a typical contact system of W UMa-type with the secondary radius of
about 0.5 $R_{\sun}$, yet the detailed observations of this binary obtained
by \citet{ruc15} indicate a much more complex geometry of the system, where
both components do not actually fill their Roche lobes but are
surrounded by the equatorial flow extending up to the inner critical
surface. The existence of such a flow in contact and near contact
binaries was predicted by \citet{ste09} and \citet{stekir13}.  The
flow redistributes energy between the components resulting in an
apparent equality of their effective temperatures what we also see in
V8 (see Table~\ref{data}).

The observational parameters of V8 result from the solution of the
light curve only. No radial velocity curve exists for the variable
which makes the solution rather uncertain. Fortunately, one of the
eclipses is total so the range of likely parameter values is not
too large. The best fitting model has the values of the main
parameters satisfactorily close to the observed ones (see
Table~\ref{data} and Fig.\ref{v8}) but the radius and temperature of
the accretor are significantly lower than observed. This results most
likely from the fact that the donor transferred to the accretor not
only its outer layers with the original hydrogen content but also some
amount of matter enriched in helium by the nuclear reactions. As a
result, the accretor behaves like a more evolutionarily advanced star,
compared to the model. Our simple modeling procedure does not take
this into account. It assumes that all the transferred matter is
hydrogen rich so past mass transfer the accretor has the hydrogen
content very close to a zero age MS star. 

The discrepancy is closely
related to another caveat of the best fitting model. To reproduce
satisfactorily the observed parameters it was necessary to assume an
initial mass ratio $q_{\mathrm{init}} =
M_{1,{\mathrm{init}}}/M_{2,{\mathrm{init}}} = 4.47$. During the rapid
mass transfer after RLOF the orbital period decreased to about 0.1 d,
which resulted in the overflow of the outer critical Roche surface by
the common stellar surface. The assumption of the conservative mass
transfer breaks down very likely in this case However, to allow for
the nonconservative mass transfer, two additional free parameters
describing the amount of mass and AM lost from the system would be
necessary. Because we avoid this, the conservative mass transfer can
be considered as a limiting case when all the matter from the common
envelope ultimately returns to the stars. Note, that the model of V8
is the only one with such an extreme initial mass ratio.

\section{Discussion}
\label{sect:disc}

We have reproduced the present configurations of the considered systems assuming 
conservative evolution of isolated binaries (apart from the mass and AM 
loss by stellar winds). Obviously, this result does not prove that V209, 
V60, V228 and V8 are formed in this way -- it just indicates such a
possibility. Approximating the evolution of binary components with single 
star models works well only in case of V209. While for the remaining three BSs
orbital periods and component masses also agree with observations, radii and
temperatures derived from modeling differ from the observed ones by several 
standard deviations, which points to the limitations of the simple evolutionary 
code employed in this paper. 
A sequence of evolutionary models of
  single stars describes correctly the evolution of close
  binary components only till the rapid mass exchange. Later on, the
  evolution of both components deviates increasingly from that of
  single stars, in particular when the accretor begins to receive nuclearly 
  processed matter from the deeper layers of the donor. 
 An obvious improvement 
  would be to replace the present code with a binary evolution code solving the
  equations of the internal structure of each component at every time
  step and allowing for the mass transfer between the components. The next step
  would be to incorporate the dynamical processes which take place in close
  binaries but are still poorly understood, like the rapid mass transfer following RLOF with an
  unknown amount of mass and AM lost from the system \citep{sarna96}, or the mass flow
  carrying thermal energy from the primary to secondary component
  in a contact binary \citep{ste09, ruc15}. 

The BS progenitors identified in the present survey are close binaries.
Below we argue that this result is calculationally robust, and physically 
feasible. 

Uncertainties of the evolutionary modeling procedure outlined
in Section \ref{sect:code} are discussed in detail by \citet{stekir15}. 
Here we only make use of their most important conclusions. 
  The progenitors of V209, V60, V228 and V8 have lost, respectively, $\Delta M$ 
  = 7.7\%, 8.8\%, 9.7\% and 7.8\%, and $\Delta$AM = 67\%, 53\%, 72\% and 71\%.
  An increase of the mass loss rate by a factor of 2 entails 
  an increase of $M$ by $\Delta M$, whereas its decrease by 50\% entails a 
  decrease of $M$ by $\Delta M$/2. Similarly, decreasing/increasing
 the AM loss rate 
  by 30\% requires a decrease/increase of the initial AM by the same percentage 
  of $\Delta$AM, corresponding to initial orbital periods of 0.84 and 2.88 d for 
  V209, 1.30 and 3.03 d for V60, 1.63 and 2.71 d for V228, and 2.82 and 3.71 d 
  for V8 (calculated for $M$ kept constant).   

Thus, varying the coefficients in Equations (7) and (8) within observationally 
acceptable limits cannot cause the model progenitors to leave the short-period
regime.

The next question to ask is: can close binaries be sufficiently abundant 
in a GC to account for the observed binary BSs? One can mention in this 
respect efficient tightening of binaries in triple systems by the Kozai 
mechanism. The presence of a third body stimulates the so called Kozai cycles
which, together with the tidal energy dissipation, can shorten an orbital 
period of 1-2 weeks down to a couple of days within less than $10^8$ years
\citep{kis06,fab07,nao14}. If this is the principal source of close binaries 
in GCs, then a local maximum around 2-4 d may be expected in the period 
distribution of young cluster binaries, similar to that observed by \citet{tok02} 
among spectroscopic subsystems of visual multiple stars. Close binaries can
also be formed due to so called ``hardening'' encounters with single or binary 
cluster members \citep[see for example][and references 
therein]{leigh15}. Note that the progenitors do not have to be formed as 
close binaries -- it is sufficient for them to tighten shortly before 
the more massive component leaves the main sequence, so that there is ample
time for the hardening to occur. However, the recent research of \citet{hyp16} 
suggests that most of the evolutionary BSs result from the unperturbed 
evolution of primordial binaries.

Whether primordial or hardened, 
all the four binaries considered here have experienced a rapid mass exchange 
with the mass ratio reversal. Their present primaries are main-sequence objects 
substantially enriched in matter from the companion. An exception is V60
whose primary is still accreting, and may be quite far from thermal equilibrium
\citep{mnr13}. The secondaries are evolutionarily advanced objects
with considerable helium cores. Two of them are already heading
towards the sdB domain, and the third one (in V228) should do it
soon. The most probable scenario for the ultimate fate of all the four
systems involves the common envelope (CE) phase beginning when the present primaries 
reach the RGB, and followed either by a merger \citep{tyl11,stekir15} or the
formation of a close binary with two degenerate stars.

The good agreement between the ages of V209 and V8 and the ages of their
parent clusters does not exclude the possibility that they were formed 
after the bulk of the cluster. Assuming e.\thinspace
g. that V209 should belong to a younger subpopulation, it is possible
to slightly increase the initial mass of the donor and to accordingly 
modify the mass of the other component together with the initial orbital 
period, so that the evolutionary time scale of the binary will be
shorter. Similarly, the age of model (4) describing V8 can be altered
by an appropriate modification of the donor mass as is demonstrated in
Figs.\ref{m0.80}-\ref{m0.90}.  

Binaries V60 and V228 are apparently younger than M55 and 47 Tuc or, in other
words, are ``overmassive'' for the ages of their parent clusters. While 
it is tempting to suggest that they are members of younger subpopulations, 
it would be premature to do so. First, it is conceivable that overmassive 
systems are products of multiple mergers or collisions \citep{sand03,leigh11}. 
Second, as we argue in the following paragraphs, available stellar evolution 
models are simply not 
reliable enough to draw such far-reaching conclusions. In particular, an apparent 
age discrepancy may emerge whenever MS life-times of models applied to 
determine the age of the cluster differ from those of models applied to 
evolve individual binaries. We note, however, that the cases of V60 and V228
are not unique. Based consistently on Dartmouth evolutionary models, 
\citet{kal15} found that the detached main sequence binaries V40 and V41 in
NGC 6362 differ in age by $1.3\pm0.4$ Gyr, i.e. statistically significant
at the 3$\sigma$ level. In our opinion, all these discrepancies are vexing
enough to be closely examined by more accurate observations and better modeling.

A broader context of our findings is origin and evolution of GCs. 
Obtaining the absolute age of a cluster involves many steps in transformation 
of observational data into theoretical parameters to which models can be fit.
Both the poor data and oversimplified theory may generate incoherent results or 
even paradoxes, like cluster ages significantly exceeding the Hubble time. The
latter values were still being derived in late 90's, as emphasized by \citet{van00}, who 
thoroughly rediscussed the issue and obtained ages between 8 and 14 Gyr. While 
the basic paradox had been removed, the problem is far from settled. Although
new methods of age determination have been introduced in addition to the original 
isochrone fitting, all of them remain strongly model-dependent, be it  
sophisticated isochrone construction \citep{dot16} or comparing absolute 
parameters of turn-off stars or detached binaries with the output of stellar 
evolution codes \citep[][project CASE, see Introduction]{marf09}. Moreover, 
in spite of constant progress in modeling significant differences exist among 
codes used by different authors. Important physical processes like convection, 
semiconvection, diffusion, mixing or rotation are usually described by single-value
parameters adopted individually in each code. Cool, solar type stars generate winds 
induced by their magnetic activity, but the resulting mass and AM loss has been largely 
neglected when computing evolutionary tracks. Only very recently some codes, like 
MESA, have been including such losses as an option.  

Even Y and Z -- the fundamental parameters describing the chemical composition of 
a star -- do not always have a well-defined meaning, since the [Fe/H] index which is often 
used as an input parameter may relate to different values of Z$_{\sun}$. Moreover,
different transformations [Fe/H]$\rightarrow$Z may be used, and different correlations 
between Z and Y may be assumed. No wonder that the resulting stellar parameters differ
from one code to another. The codes are typically calibrated on the
Sun, so they reproduce MS solar type stars in a consistent way. Yet,
the most important parameter for the correct determination of 
GC ages i.\thinspace e. MS life time can differ for some stars
quite considerably. To give an example, the old Padova model of a 0.8
$M_{\sun}$ star with Z=0.001 has MS life time equal to 13.5 Gyr
\citep{gir00}, whereas the new PARSEC model of the same star only 
11.95 Gyr \citep{bre12}. Another example: a recently
published by \citet{lag12} 0.85 $M_{\sun}$ model with Z=0.0001 has the
MS life of 11.14 Gyr, as compared to 9 Gyr for the PARSEC
model with the same Z. Differences in time scales between models beyond 
the MS are still greater. This diversity became recently of concern for 
a number of authors who discuss the influence of several physical
mechanisms on models produced by different codes and compare them to
well calibrated observational data \citep{nat12,mar13,ghe15,mey15,
stan15,val16a,val16b}. However, uncertainties of stellar evolution models 
still remain the primary factor impeding the accurate determination of the 
absolute ages of globular clusters. Observational uncertainties will 
dramatically decrease when Gaia parallaxes become available \citep[distances to many 
GCs will be known with an accuracy better than one per cent;][]{pan13}.
If the theory is to catch up with observations, much more effort in exposing and 
diagnosing its problems is needed. We believe that evolutionary simulations of binary 
blue stragglers is a very useful tool to achieve this goal.  

\section{Conclusions}
\label{sect:summ}

   \begin{enumerate}
      \item The results of the present investigation demonstrate an
        importance of the close binary evolution in the formation of
        blue stragglers. In particular, it was shown that blue
        stragglers result from rejuvenation of an initially less
        massive component fed by the hydrogen rich matter from its
        companion overflowing the inner Roche lobe.
      \item To reproduce the present parameters of V209 in $\omega$ Cen it was
        necessary to adopt the increased helium content Y = 0.40 of
        this binary. This is further evidence in favor of the 
        existence of the helium-rich subpopulation in this cluster.
      \item The best fitting models of V60 in M55 V228 in 47 Tuc 
        are too massive for the age of the parent cluster as it is currently
        estimated. Unless this is an apparent effect resulting from an
          inconsistency between models used for GC age determination
          and binary modeling, these systems may belong to younger 
        subpopulations or
        may result from dynamical multibody interactions. 
      \item Different evolutionary codes produce mutually inconsistent
        stellar parameters. Better description of physical processes
        influencing the stellar evolution is urgently needed. Until
        this shortage is overcome, the absolute age determination of
        globular clusters will be plagued with substantial
        uncertainties. 
      \item Studying the evolution of binary BSs in GCs   
        emerges as a new research 
        tool which provides an independent insight into interactions between GC 
        members, and enables sensitive tests of stellar evolution codes. 
   \end{enumerate}

\begin{acknowledgements}
This paper is dedicated to the memory of Janusz Kaluzny, founder and long-time leader 
of CASE, who prematurely passed away in March 2015.\\ 
We thank an anonymous
  referee for very insightful remarks which significantly improved the
  final shape (czy moze presentation?) of the paper. \\
wg mojej implementacji Polglish mo\.ze 
by\'c albo "improved the presentation" (full stop here), albo improved the final shape of 
the paper. Nie mam preferencji :)\\
AAP acknowledges partial support from the Polish National Science Center (NCN) grant
2015/17/B/ST9/02082. MR was supported by the NCN grant DEC-2012/05/B/ST9/03931.
\end{acknowledgements} 

\bibliographystyle{aa}

\end{document}